# Improving the Efficiency of OpenCL Kernels through Pipes

Mostafa Eghbali Zarch
*North Carolina State University*
Raleigh, NC, USA
meghbal@ncsu.edu

Michela Becchi
*North Carolina State University*
Raleigh, NC, USA
mbecchi@ncsu.edu

*Abstract*—In an effort to lower the barrier to the adoption of FPGAs by a broader community, today major FPGA vendors offer compiler toolchains for OpenCL code. While using these toolchain allows porting existing code to FPGAs, ensuring performance portability across devices (i.e., CPUs, GPUs and FPGAs) is not a trivial task. This is in part due to the different hardware characteristics of these devices, including the nature of the hardware parallelism and the memory bandwidth they offer. In particular, global memory accesses are known to be one of the main performance bottlenecks for OpenCL kernels deployed on FPGA.

In this paper, we investigate the use of pipes to improve memory bandwidth utilization and performance of OpenCL kernels running on FPGA. This is done by separating the global memory accesses from the computation, enabling better use of the load units required to access global memory. We perform experiments on a set of broadly used benchmark applications with various compute and memory access patterns. Our experiments, conducted on an Intel Arria GX board, show that the proposed method is effective in improving the memory bandwidth utilization of most kernels, particularly those exhibiting irregular memory access patterns. This, in turn, leads to performance improvements, in some cases significant.

## I. INTRODUCTION

Over the past several years, there has been an increasing trend toward using heterogeneous hardware in single machines and large-scale computing clusters. This trend has been driven by demands for high performance and energy efficiency. Initially, heterogeneity has mostly involved the use of GPUs and Intel many-core processors alongside multi-core CPUs [1]. More recently, due to their compute capabilities and energy efficiency, the trend has evolved to include Field Programmable Gate Arrays (FPGAs) [2] in high-performance computing clusters and data centers. Today, Microsoft Azure and Amazon Web Services include FPGA devices in their compute instances [3] [4].

Hardware heterogeneity involves significant programmability challenges. Without a unified programming interface, not only are users required to become familiar with multiple programming frameworks, but they also need to understand how to optimize their code to various hardware architectures. To address this challenge, the Khronos group has introduced a unified programming standard called OpenCL, which is intended for accelerated programming across different architectures [5]. This programming model initially targeted CPUs and GPUs. At the same time, programming FPGAs using low-level hardware description languages (HDLs) has traditionally been considered a specialized skill. To facilitate the adoption of FPGAs, vendors have spent substantial resources on the design and the development of OpenCL-to-FPGA toolchains, including runtime libraries and compilers allowing the deployment of OpenCL code on FPGA. Intel and Xilinx, two major FPGA vendors, are now providing their own OpenCL-to-FPGA development toolchain and runtime system [6] [7].

Although OpenCL increases portability and productivity, there is often a significant performance gap between an OpenCL and a hand-optimized HDL version of the same application [8]. Bridging this performance gap while limiting the development effort requires exploring existing OpenCL-to-FPGA optimizations and designing new ones. Several papers have aimed to improve the efficiency of existing OpenCL code (often tailored to GPUs) on FPGA through platform-agnostic and specific compiler optimizations and scheduling techniques [9] [10] [11] [12].

Performance portability is one of the major issues when using OpenCL-to-FPGAs SDKs, especially for applications originally encoded for a different device (e.g., a GPU). It has been shown that OpenCL code tailored to one platform often performs poorly on a different platform [13]. The origin of the performance portability issues between GPUs and FPGAs lies in the different architectural characteristics of these two platforms. Specifically, there are three fundamental factors that affect the performance portability between GPUs and FPGAs. First, the form of parallelism that these devices offer. FPGAs leverage deep pipelines to exploit parallelism across OpenCL work-items, while GPUs rely on concurrent, SIMD execution of threads (or work-items). Second, the off-chip memory bandwidth of current FPGA boards is much lower than that offered by high-end GPUs, which results in inefficient memory operations and lower overall application performance. Third, while GPUs provide relatively efficient support for synchronization primitives like barriers and atomic operations, barriers on FPGAs result in a full pipeline flush, leading to significant performance degradation.

In this work, we explore and evaluate the use of the feed-forward design model to improve the performance of OpenCL code on FPGA. The proposed model splits each kernel into two kernels - a *memory kernel* and a *compute kernel* - connected through pipes. At a high level, the model aims to increase the memory bandwidth utilization, reduce the memory units' congestion, and maximize the instructions concurrency within the application. We show that the feed-forward design model allows the offline compiler to generate designs with more efficient memory units and increased instruction parallelism, leading to better performance with a low resource utilization overhead. A simplified version of this scheme has been explored in [14] on simple micro-kernels, in most cases leading to performance degradation over the original single work-item version of the code. In this work we show that, when generalized and applied to more complex and less regular kernels, this technique can achieve up to a $65\times$ speedup over the single work-item version of the code, and an average $20\times$ speedup across a set of diverse applications from popular benchmark suites [15] [16].

Our exploration is structured as follows. First, based on recommendations from Intel's OpenCL-to-FPGA documentation [17], we convert SIMD-friendly code into serial code (i.e., a single work-item kernel). Second, using the the feed-forward design model, we split each kernel into two kernels (*memory* and *compute* kernels), thus separating global memory reads from the rest of the instructions inside the kernel. In order to minimize the data communication latency, we connect these two kernels through pipes. Lastly, we explore increasing the concurrency by having multiple versions of *memory* and *compute* kernels working on different portions of the data. In our experiments, we first compare the performance and resource utilization of the original kernels and the versions using the feed-forward design model. Then, we analyze the impact of the feed-forward model on other best-practice optimizations.

In summary, this work makes the following contributions:

- Proposing a systematic code transformation method to apply the feed-forward design model to existing OpenCL kernels;
- Studying benefits and limitations of the feed-forward design model;
- Exploring optimizations enabled by the feed-forward design model;
- Evaluating the use of our method on a set of benchmark applications and microbenchmarks with various memory access and compute patterns. Our experiments show performance improvements from our method from 30% up to $86\times$ across benchmarks at the cost of a modest resource utilization overhead.

## II. BACKGROUND

### A. OpenCL for FPGA

OpenCL allows programmers to write platform-agnostic programs and deploy them on a wide range of OpenCL compatible devices. An OpenCL application consists of two types of code: host code and device code. The host code is responsible for data allocation on the host machine and accelerators (devices), communication setup and data transfer between host and devices, configuration of the accelerators, and launching the device code on them. The device code contains the core compute kernels, is written to execute on one or multiple platforms, and is often parallelized. In OpenCL terminology, a kernel consists of multiple *work-items* evenly grouped in *work-groups*. When deployed on GPU, work-items correspond to threads, and work-groups to thread-blocks.

OpenCL kernels can be in two forms: *NDRange* or *single work-item*. NDRange kernels consist of multiple work-items, distinguishable through their local and global identifiers, launched by the host code for parallel execution. This model is widely used for programming CPUs and GPUs; on FPGAs, concurrent execution of work-items is enabled through pipeline parallelism. Single work-item kernels have a serial structure, with only one work-item launched by the host code. The single work-item model is preferred when the NDRange version of the kernel presents fine-grained data sharing among work-items. Single work-item kernels are often recommended by FPGA vendors [7], partially because writing the same kernels in NDRange fashion might require expensive atomic operations or synchronization mechanisms to ensure correctness. In cases that the OpenCL application is in NDRange form like the baseline implementation of the benchmarks in [16] and [15], programmers can construct the single work-item version by embedding the body of the NDRange baseline kernel within a nested loop. The outer and inner loops must have the work-group and work-item sizes as the loop iteration count, respectively.

Major FPGA vendors - such as Intel and AMD/Xilinx - currently provide OpenCL-to-FPGA SDKs to facilitate FPGA adoption by a wide range of programmers with different skills. However, the automatic generation of FPGA code often incurs performance portability issues, especially when the OpenCL code was originally optimized for a different device, such as a GPU. To bridge the performance gap between FPGAs and other devices, it is critical to understand the performance limiting factors on FPGA, and design FPGA specific optimizations.

### B. OpenCL memory model

The OpenCL memory model, shown in Figure 1, includes five sections. The *host's global memory* is directly accessible by the host processor; data in it can be then transferred to the device. The *device's global memory* is accessible by both the host processor and all the work-items of the kernels run on the device. The *constant memory* can be read and written by the host code but only read by the device kernels. Physically, the device's global and constant memories are usually mapped to memory chips connected to the FPGA device. However, in some cases, they are mapped to distributed memories located within the FPGA fabric [6], [13]. The last two sections are high-throughput and low-latency memory regions known as *local* and *private memories*. The former region is shared and accessible by all the work-items within a single work-group. The latter region is only visible to a single work-item. These two memory regions are often implemented using block RAMs or registers in the FPGA fabric.

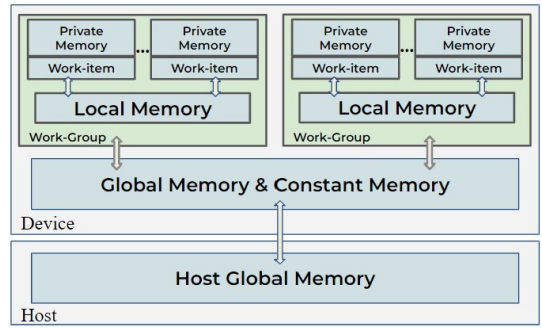

Fig. 1: OpenCL memory model

### C. Use of Load/store units on FPGA

In order to understand the effect of compiler optimizations and scheduling techniques on memory operations, it is essential to know how OpenCL-to-FPGA compilers implement memory operations using load/store units (LSUs). In the rest of the paper, we refer to Intel's OpenCL-to-FPGA SDK as the offline compiler. The offline compiler can instantiate several types of LSUs depending on the inferred memory access patterns of the memory operations. Two pieces of information used to determine the LSU type to be used are the memory region accessed (i.e., global versus local memory) and the types of LSUs available on the target FPGA platform. There are three LSU types available to the Intel's offline compiler: *burst coalesced*, *prefetching*, and *pipelined* LSUs. Burst coalesced LSUs are often used as the default type. This type of LSU is the most resource-hungry memory module, and it is designed to buffer memory requests until the largest possible burst of data read/write requests can be sent to the global memory. Prefetching LSUs leverage a FIFO to read large blocks of data from global memory and aim to keep the buffer full of valid data. This type best fits memory operations with a sequential memory access pattern. For local memory accesses, the offline compiler typically instantiates pipelined LSUs, which submit memory requests in a pipeline manner as soon as they are received. In some cases, the offline compiler uses pipelined LSUs as an alternative for global memory accesses, resulting in slower but more resource-efficient memory units.

### D. OpenCL pipes and channels

**Pipes** - OpenCL applications consisting of multiple kernels require efficient mechanisms for inter-kernel communication. Using global memory for this purpose requires race-free global memory accesses or the use of atomic operations and barriers, which can be inefficient on FPGA. The OpenCL standard provides a mechanism to pass data between kernels, called "pipes". Essentially, pipes represent ordered sequences of data items. Each pipe has separate write and read endpoints, allowing an OpenCL kernel to write to one endpoint of the pipe while another kernel reads from the other endpoint. By allowing concurrent execution of interconnected kernels, pipes enable pipeline parallelism across kernels. It is worth mentioning that, in OpenCL, the host and device(s) can also communicate through pipes, a feature not used in this paper.

**Channels** - Intel provides an OpenCL extension called "channels" as a mechanism for data communication between kernels [17]. Channels allow concurrently running kernels to communicate without the involvement of the host processor or the use of device's global memory, and provide a mechanism for efficient inter-kernel synchronization. Programmers can define the depth of the channels as an input attribute. The offline compiler considers this input to be the minimum depth of that specific channel, and may increase the depth of the channels in two situations: first, if there is a need to balance the re-converging paths through multiple kernels; and second, to achieve better loop pipelining [17]. There are two kinds of channels: *blocking* and *non-blocking*. A blocking channel operation (read/write) will stall the kernel until the operation returns successfully. In contrast, non-blocking channel operations return the results of the operation as successful or failed immediately through an additional flag.

### III. IMPLEMENTING THE FEED-FORWARD DESIGN MODEL

In this section, we first motivate our work (Section III-A). Then, we provide a high level view of the feed-forward design model on FPGA and discuss limitations to its applicability (Section III-B). At last, we present our approach for systematically transforming a generic kernel to use this design model, we elaborate on strengths and weaknesses of the method, and we propose some optimizations (Section III-C).

### A. Motivation

Global memory accesses are known to be one of the main performance bottlenecks for OpenCL kernels implemented on FPGAs. Wang et al. [18] measured the memory bandwidth of sequential and random memory accesses for different variable types and concluded that the presence of random memory accesses within a kernel can limit the memory bandwidth achieved drastically. In addition, they observed that severe lock and memory bandwidth overhead limit the throughput of many of the OpenCL kernels they considered in their analysis. Optimizing memory accesses in OpenCL code is not a trivial task. While Intel's SDK gives the programmer some level of control over the type of load/store units (LSUs) instantiated by the offline compiler to handle memory instructions, selecting the optimal LSU for each memory operation requires good understanding of the hardware and insights on the offline compiler's operation.

By studying the memory analysis reports of the offline compiler for a set of real and synthetic OpenCL kernels with various memory access patterns, we identified two significant factors affecting the kernel's performance: (i) the type and configuration of the LSUs instantiated by the compiler to handle global memory instructions, and (ii) the presence of loop-carried dependencies on global memory variables. The offline compiler associates to each loop an *initiation interval* (II), which represents the number of clock cycles between the launch of successive loop iterations. In the presence of a loop that contains loop-carried dependencies, the offline compiler serializes its execution, resulting in a high initiation interval and, consequently, low throughput. In our study, we noticed that in many applications, the offline compiler identifies loop-carried dependencies even in situations where the algorithm does not imply one. In the rest of the discussion, we refer to these dependencies as *false* loop-carried dependencies. This conservative approach can drastically degrade the performance as the selection of load/store units and the initiation interval of the loops will be affected.

In this work, we study the use of the feed-forward design model on FPGA, and we propose a systematic approach to transform generic OpenCL kernels to use this model. This design model implicitly exposes information on the characteristics of the memory operations and data dependencies within an OpenCL code to the offline compiler. Importantly, the use of the feed-forward design model implies the absence of true loop-carried dependencies between load and store instructions on global memory. Moreover, it enables increasing the concurrency among memory instructions using multiple load units. In turn, this technique can result in synchronization-free kernels with high memory bandwidth utilization.

### B. The feed-forward design model on FPGA

In the feed-forward design model, the computation is broken down into two kernels: a *memory kernel* and a *compute kernel*. The former is responsible for loading values from global memory, and the latter for performing computation on the loaded data. In order to allow for efficient implementation, the two kernels should be connected through a hardware mechanism that does not involve the use of global memory, allowing the second kernel to avoid global memory loads. As explained in Section II, programmers can use *pipes/channels* to establish this communication.

Previous work has explored the use of channels to connect multiple kernels that are already part of an application, creating efficient kernel pipelines [19]. Our study uses channels for a different purpose, and targets also applications that consist of a single kernel. Specifically, our proposed transformation *splits* existing kernels to enable the use of the feed-forward model, and then uses channels to connect the generated sub-kernels. Other work has explored the use of channels on simple hand-written kernels, with the goal of improving the efficiency of their memory operations [14] [18]. However, the performance advantages reported are limited, partially due to the simplicity and the regular memory access patterns of those kernels. Our goal is to propose a general code transformation method that allows applying the feed-forward execution model to diverse OpenCL kernels with regular or irregular memory access patterns, and with simple or complex control flows.

While the use of the feed-forward design model can improve the memory bandwidth utilization of OpenCL code deployed on FPGA, there are limitations to its applicability in case of iterative applications. Specifically, the presence of loop-carried dependencies can prevent or affect the application of this design model to OpenCL applications. This shortcoming is due to the lack of a well-defined technique to introduce global inter-kernel synchronizations between concurrent OpenCL kernels running on FPGA. However, not all loop-carried dependencies are problematic, and some of them can actually be handled by the code transformation method we propose. In Section III-C, we categorize loop-carried dependencies and discuss their effect on the application of the feed-forward design model to existing OpenCL kernels.

### C. Feed-forward design transformation

Here, we first present our systematic approach to transform existing OpenCL kernels to leverage feed-forward execution. We then discuss the effect of various categories of loop-carried dependencies on the applicability and efficacy of the method. Finally, we discuss the generalization of our method to multiple memory and compute kernels.

**Code transformation method** - Without loss of generality, in Figure 2 we illustrate the method on a single work-item kernel. We note that, while our method is also applicable to NDRange kernels, Intel recommends using single work-item kernels when using channels and pipes [17]. The example in Figure 2 uses notations from the Intel OpenCL-to-FPGA SDK. Our method consists of the following steps:

① Identify instructions that read from global memory (lines 2, 4, 7, 12, 13, & 14 of the baseline kernel in Figure 2a).
② Verify that the kernel to be transformed (baseline kernel) is free from *true memory loop-carried dependencies* (true MLCDs), as defined below. Note that the list of loop-carried dependencies identified by the offline compiler is provided in the compiler's report. Unless removed, true MLCDs prevent the safe application of the method.
③ Allocate a local variable for each load instruction used in the condition of a conditional statement or loop (lines 2, 14, 15, 18, 19 in Figure 2b).
④ Copy the baseline kernel (Figure 2a) into two different kernels, namely, the *memory* kernel (Figure 2b) and *compute* kernel (Figure 2c).
⑤ Define a channel for each global load instruction using the proper data type. If the same data item is loaded repeatedly in the baseline kernel, only assign a single channel to it (as for loads in lines 13 & 14 in Figure 2a).
⑥ Add to the *memory* kernel a write-to-channel instruction for each read from global memory unless the loaded value is only used as an index by another load instruction (lines 3, 6, 12, 16, and 20 of Figure 2b). Coherently with step ⑤, if a data item is loaded repeatedly in the baseline kernel, write the corresponding channel only once.
⑦ In the *compute* kernel, replace instances of reading from global memory with a read from the assigned channel (lines 2, 5, 6, 9, and 11 in Figure 2c). If the same data item is loaded repeatedly, use the value read from the channel of the first instance (line 13 of 2c).
⑧ Remove any arithmetic operations, global memory stores, or local memory objects from the *memory kernel* (unless they affect the control flow paths or indices of load instructions).

```
1.    for (int tid = 0; tid<num_nodes; tid++){
2.       if(c_array[tid] == -1){
3.          *stop = 1;
4.          int start = row[tid];
5.          int end;
6.          if (tid + 1 < num_nodes)
7.             end = row[tid + 1] ;
8.          else
9.             end = num_edges;
10.         float min = BIGNUM;
11.         for(int edge = start; edge < end; edge++){
12.            if (c_array[col[edge]] == -1){
13.               if(node_value[col[edge]] < min)
14.                  min = node_value[col[edge]];
15.            }
16.         }
17.         min_array[tid] = min;
18.      }
19.   }
```

(a) Baseline kernel (not using feed-forward execution)

```
1.    for (int tid = 0; tid<num_nodes; tid++){
2.       int c_arr = c_array[tid];
3.       write_channel_intel(ch0, c_arr);
4.       if(c_arr == -1){
5.          int start = row[tid];
6.          write_channel_intel(ch1, start);
7.          int end;
8.          if (tid + 1 < num_nodes)
9.             end = row[tid + 1] ;
10.         else
11.            end = num_edges;
12.         write_channel_intel(ch2, end);
13.         for(int edge = start; edge < end; edge++){
14.            int col_e = col[edge];
15.            int c_arr1 = c_array[col_e];
16.            write_channel_intel(ch3, c_arr1);
17.            if (c_arr1 == -1) {
18.               int col_e1 = col[edge];
19.               int node_value_c = node_value[col_e1];
20.               write_channel_intel(ch4, node_value_c);
21.            }
22.         }
23.      }
24.   }
```

(b) Resulting *memory* kernel for feed-forward execution

```
1.    for (int tid = 0; tid<num_nodes; tid++){
2.       int c_arr = read_channel_intel(c0);
3.       if(c_arr == -1){
4.          *stop = 1;
5.          int start = read_channel_intel(c1);
6.          int end = read_channel_intel(c2);
7.          float min = BIGNUM;
8.          for(int edge = start; edge < end; edge++){
9.             int c_arr1 = read_channel_intel(c3);
10.            if (c_arr1 == -1){
11.               float node_val = read_channel_intel(c4);
12.               if(node_val < min)
13.                  min = node_val;
14.            }
15.         }
16.         min_array[tid] = min;
17.      }
18.   }
```

(c) Resulting *compute* kernel for feed-forward execution

Fig. 2: Feed-forward design model example

⑨ Remove from the resulting kernels empty control flow paths and values not further used (i.e., apply a dead code elimination pass).

⑩ Instantiate *memory kernel* and *compute kernel* multiple times to increase concurrency and adjust the main loop trip counts accordingly (more details provided below).

⑪ Repeat step ⑨.

⑫ Replace the baseline kernel launch inside the host code with invocations of the *memory* and *compute* kernels on separate queues.

We note that this technique can generate a *memory kernel* with a simplified control flow graph (CFG) compared to the original (baseline) kernel. A less complex CFG results in fewer stalls for global memory reads, hence leading to a higher memory bandwidth utilization. This, in turns, can lead to better overall performance.

**Loop-carried dependencies** - Here, we discuss the benefits and limitations of our method in the presence of loop-carried dependencies (LCDs) in the kernel to be transformed. LCDs can be in the form of *memory loop-carried dependencies* (MLCDs) or *data loop-carried dependencies* (DLCDs).

Memory LCDs - A MLCD occurs when a data value stored in global memory in one iteration of a loop is requested in a later iteration of the same loop. This kind of data dependencies implies that the device needs to schedule the read and write operations serially to ensure correctness of the results. As a consequence, when a MLCD is detected, the offline compiler will serialize the execution of the corresponding loop.

Figure 3(a) shows an example of MLCD in which the statement at line 2 depends on the result of the execution of the statement at line 4 in the previous iteration. The offline compiler will serialize this loop to ensure correctness. As mentioned earlier, loop serialization generally leads to a high initiation interval, negatively affecting performance. When identifying MLCDs within a kernel, the Intel's offline compiler takes a conservative approach and flags a MLCD when it cannot determine whether a memory instruction is involved in a MLCD or not. This can result in *false* MLCDs being detected. On the other hand, the Intel's compiler does not detect MLCDs across kernels. Recall that our proposed transformation separates global memory loads (performed only in the memory kernel) from global memory stores (performed in the compute kernel), thus removing all intra-kernel MLCDs.

All LCDs are listed in the offline compiler's report. By identifying false MLCDs dependencies (either manually or through static analysis techniques), the programmer can determine when the code transformation described above is safely applicable. When applying the feed-forward design model on the Maximal Independent Set application (see Section IV), for example, removing the false MLCDs results in improving the maximum global memory bandwidth utilization from 208 MB/s to 2116 MB/s and results in a 6.35× speedup over the original kernel.

Data LCDs - A DLCD occurs when a local variable updated in one iteration of a loop is read in a later iteration. Like

```
1.    for (int tid = 1; tid<num_nodes; tid++){
2.      int a = output[tid-1];
3.      int b = input[tid];
4.      output[tid] = a + b;
5.    }
```

(a)    Memory loop carried dependency

```
1.    for (int tid = 5; tid<num_nodes; tid++){
2.      int r = 0;
3.      for (int iter=0; iter<5; iter++){
4.        int a = input[tid-iter];
5.        r += a;
6.      }
7.      output[tid] = r;
8.    }
```

(b) Data loop carried dependency baseline

```
1.    for (int tid = 5; tid<num_nodes; tid++){
2.      for (int iter=0; iter<5; iter++){
3.        int a = input[tid-iter];
4.        write_channel_intel(c0, a);
5.      }
6.    }
```

(c) Data loop carried dependency Memory kernel

```
1.    for (int tid = 5; tid<num_nodes; tid++){
2.      int r = 0;
3.      for (int iter=0; iter<5; iter++){
4.        int a = read_channel_intel(c0);
5.        r += a;
6.      }
7.      output[tid] = r;
8.    }
```

(d) Data loop carried dependency Compute kernel

Fig. 3: Loop-carried dependencies examples

MLCDs, DLCDs result in loop serialization being performed by the offline compiler. Figure 3(b) shows an example with a DLCD involving the statements at lines 3 and 5, causing the offline compiler to serialize the loop's execution. Again, serialization prevents pipeline parallelism and reduces the memory bandwidth utilization, ultimately limiting kernel performance. However, after applying our transformation to the baseline kernel, the DLCD becomes part only of the compute kernel, which is free of memory operations. This will allow the offline compiler to schedule load instructions in the memory kernel more efficiently, since now these instructions are in a loop with no DLCDs. Figure 3(c) and 3(d) show how the DLCD in the loop in Figure 3(b) is left only in the compute kernel, allowing the offline compiler to schedule the memory instructions in memory kernel in a pipelined manner.

To summarize, our approach can benefit kernels with LCDs in two ways. First, in the presence of DLCDs in the original kernel, the transformation causes the DLCD to be present only in the compute kernel, allowing the memory kernel to be pipelined. Second, in the absence of true MLCDs, the transformation enables loop pipelining (even if the original kernel contains false MLCDs).

**Enabling feed-forward design model with multiple memory and compute kernels** - The most significant advantage of our proposed technique is enabling the feed-forward design model to increase the memory bandwidth utilization for load instructions. More performance advantages can be achieved by increasing concurrency among memory operations. In the feed-forward design model, data move from the memory kernel to the compute kernel in one or multiple words. For each word written on a pipe by the memory kernel, the compute kernel will process the data and free up the memory space assigned to the channel. Having multiple memory and compute kernels can potentially increase the global memory bandwidth achieved by the application, increasing performance at a limited resource utilization overhead.

Having multiple memory and compute kernels requires making various decisions regarding their number, the load balancing mechanism, and the buffer management scheme to be adopted. Kernel replication adds concurrency while increasing complexity and resource utilization. Intel recommends limiting the number of channels used in the design, as they can add complexity and limit overall performance. In addition, having a large number of kernels reading data from global memory concurrently can increase memory congestion and result in poor global memory bandwidth utilization. In our experiments, we did not find significant performance improvements beyond two memory and two compute kernels. Moreover, we explored using a single memory kernel and multiple compute kernels and vice versa. Our results indicate that having separate memory and compute kernels communicating directly with each other yields higher concurrency compared to having one memory send data to multiple compute kernels. We will elaborate more on these experiments in Section IV.

Different load balancing mechanisms can be used to distribute the work across multiple memory/compute kernels. These mechanisms can be classified as either static or dynamic. Unlike dynamic mechanisms, static load balancing schemes do not take into account the state of the system when making decisions. Many dynamic load balancing schemes require busy-wait or feedback mechanism implementations involving kernels polling on non-blocking channels. This form of busy-wait can result in performance degradation on FPGA. In this work, we use static load balancing to connect memory and compute kernels. When using multiple memory/compute kernels, we chose to partition the data in continuous chunks. This results in optimized load units for load instructions with regular memory access patterns.

## IV. EXPERIMENTAL EVALUATION

### A. Experimental Setup

**Hardware and Software Setup** - We ran our experiments on an Intel programmable acceleration card (PAC) with an Arria® GX FPGA. This board contains two 4 GB DDR-4 SDRAMs memory banks with a maximum bandwidth of 34.1 GB/s, and 128 MB of flash memory. This FPGA includes 65.7 Mb of on-chip memory, 1150k logic elements (ALUTs), and 3036 digital signal processing (DSP) blocks. On the host side, the machine is equipped with an Intel Xeon® CPU model E5-1607 v4 with a maximum clock frequency of 3.1 GHz.

TABLE I: Benchmarks applications and datasets

| Suite | Benchmark | Dwarves | Memory Access Pattern | Dataset |
|---|---|---|---|---|
| Rodinia | Breadth-First Search (BFS) | Graph Traversal | Irregular | #nodes=2M |
| | Hotspot (HS) | Structured Grid | Regular | Size=8192 |
| | k-Nearest Naighbors (NN) | Dense Linear Algebra | Regular | Size=8.3M |
| | Hotspot 3D (HS-3D) | Structured Grid | Regular | Size=8192 |
| | Needleman-Wunsch (NW) | Dynamic Programming | Regular | Size=8192 |
| | Back Propogation (BP) | Unstructured Grid | Regular | Size=12.8M |
| Pannotia | Floyd-Warshall (FW) | Graph Traversal | Irregular | Size=512 |
| | Maximal Independent Set (MIS) | Graph Traversal | Irregular | Size=1.58M |
| | Page Rank (PR) | Graph Traversal | Irregular | Size=1.58M |
| | Graph Coloring (GC) | Graph Traversal | Irregular | Size=1.58M |

We used the Intel FPGA SDK for OpenCL version 19.4 and Ubuntu 18.04.6 LTS.

**Benchmarks** - We performed two sets of experiments. In the first, we evaluated our method on widely used open-source benchmarks from Rodinia [15] and Pannotia [16]. These benchmark suites contain applications from different domains, and have been used in previous work on OpenCL for FPGA [12] [20] [13] [19]. Table I summarizes the main characteristics of these applications. In the second set of experiments, we used automatically generated microbenchmark kernels to evaluate the impact of the kernel characteristics (namely, memory access patterns and branch divergence) on the efficacy of the proposed method.

**Performance Metrics** - For performance, we report the speedup over the original single work-item version of each benchmark. As we mentioned in Section II Intel recommends using single work-item programming model when designing kernels with channels. For resource utilization, we report the logic utilization and use of block RAMs (BRAMs) of each implementation. The logic utilization represents an estimate of the number of half adaptive logic modules (ALMs) used by the compiler to deploy the OpenCL kernels on FPGA. ALMs are hardware logic blocks. The simplest version of ALMs contains a lookup table (LUT) and a register. The compiler reports the logic utilization as a percentage of the total number of half ALMs on the FPGA board.

### B. Experimental Results

*1) Experiments on Benchmark Applications:*
**Single Producer-Single Consumer** - Table II shows the performance impact of applying the feed-forward design model to the considered benchmark applications when using a single producer and a single consumer kernels. In all cases, we used the baseline code without any optimizations in order to isolate the impact of the feed-forward model on performance. We report the best results from running the same experiments using channels with three depths: 1, 100, and 1000. We recall that the depth parameter passed to the offline compiler indicates the minimum depth of that specific channel, but the compiler might increase the depth to improve efficiency. Our experiments showed that the channel depth parameter's setting does not affect performance significantly, proving that the Intel compiler does a good job of adjusting the channel depth.

As shown in Table II, among the benchmarks we explored, BFS, FW, BP, MIS, and NW benefit significantly from applying the feed-forward model. For all these benchmarks, the main driver for the speedup is removing false loop carried dependencies on variables located in the device's global memory.

For FW, the false loop-carried dependencies detected by the offline compiler resulted in a large initiation interval (II) of 285 for the main loop inside the kernel. We recall that the II indicates the number of clock cycles between two consecutive loop iterations, and is an indication of the efficiency of the hardware pipeline generated by the compiler. As often done on global memory loads, the offline compiler used the burst coalesced LSU type to implement load instructions, resulting in low memory bandwidth for instructions with regular memory access patterns. The use of the feed-forward design model had two effects. First, resolving those false loop-carried dependencies allowed converting the main loop to a fully pipelined loop with an II of 1. Second, it enabled the offline compiler to use a prefetching LSU for one of the three global load operations with a regular memory access pattern and increased the maximum global memory bandwidth of the kernel from 630 MB/s to 3130 MB/s. These changes resulted in a speedup up to $65\times$ compared to the baseline kernel.

BP benefits from the feed-forward implementation in a similar way. In the original kernel, the performance is limited by the main loop, which exhibits an II of 416. In the feed-forward version of the kernel, the same loop is fully pipelined, with an II of one. This decrease in the II also increased the maximum global memory bandwidth used by the kernel and resulted in a significant speedup of $44\times$ over the baseline version of the kernel. Similar trends were observed on the other three benchmarks (BFS, MIS, and NW) benefiting from the feed-forward model.

We should note that the baseline version of NW carries a true MLCD inside the main loop of the kernel. However, this loop-carried dependency is entangled to a load instruction in iteration *K* which is dependent on a store instruction in iteration *K-1*. In this case, this loop-carried dependency can be resolved in the baseline kernel using a local variable in the private memory of the device. Storing the dependency value at the end of each iteration can remove the existing loop-carried dependency. As a result, the kernel can read the same value at the beginning of the next iteration (except the first iteration). Adding this private variable results in a single work-item baseline kernel with no MLCDs. Then, applying the feed-forward design model allows achieving a 50× speedup by decreasing the II of the main loop and increasing the global memory bandwidth of the kernel.

TABLE II: Resource utilization and throughput comparison of the the feed-forward design model with the baseline

| Benchmark | Baseline Execution Time (ms) | Feed-foward Design Speedup over Baseline | Basline Logic Utilization (%) | Feed-Forward Design Logic Utilization (%) | Baseline BRAM | Feed-Forward Design BRAM |
|---|---|---|---|---|---|---|
| BFS | 6422 | 13.84 | 18.62 | 20.00 | 578 | 596 |
| HS | 22553 | 0.85 | 16.14 | 17.25 | 517 | 522 |
| NN | 18.27 | 1.03 | 16.04 | 16.46 | 376 | 376 |
| HS-3D | 31967 | 0.88 | 16.45 | 17.95 | 542 | 536 |
| NW | 26036 | 50.95 | 16.10 | 18.86 | 506 | 407 |
| BP | 140601 | 44.54 | 24.67 | 26.68 | 674 | 646 |
| FW | 41760 | 64.95 | 16.21 | 16.47 | 482 | 465 |
| MIS | 2166 | 6.47 | 21.77 | 24.44 | 803 | 807 |
| PR | 8430 | 0.96 | 20.43 | 22.52 | 703 | 709 |
| GC | 453 | 1.02 | 17.78 | 19.48 | 651 | 656 |

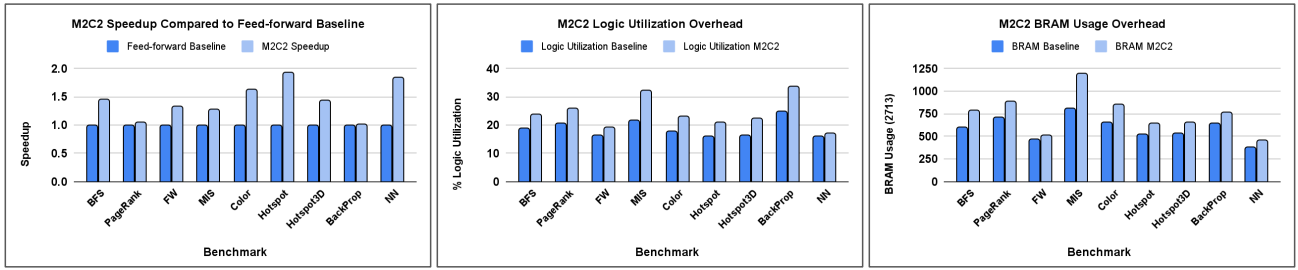

Fig. 4: M2C2 speedup and resource utilization overhead compared to the feed-forward design

The feed-forward model reports minor performance improvements on NN and GC, and small performance degradation on HS, HS-3D, and PR (see Table II). For these applications, the Intel profiler's report reveals the presence of one or multiple LSUs with high "occupancy". The LSU occupancy is an indicator of the percentage of the execution time where the LSU is not stalled. A high LSU occupancy (i.e., low number of memory stalls) is desirable, and indicates that the kernel is already making good use of the memory bandwidth and its memory accesses do not need further optimization.

As reported in Table II, our code transformation method introduces only a modest resource utilization overhead. On average, our method only increases the logic utilization by 9% and even has positive effect on BRAM usage on some benchmarks (at the expense of using more registers). The profiling and resource utilization reports from the offline compiler show that our method can optimize the load units from both the throughput and resource utilization point of view. This limits the overhead associated with transforming one kernel to two kernels. Moreover, channels are implemented using on-chip memory, hence, they introduce a very small overhead on resource utilization.

**Multiple Producers/Consumers** - While the feed-forward design model enables removing false MLCDs and increasing the maximum global memory bandwidth of the kernel, it also enables using multiple memory and compute kernels to increase the concurrency among the instructions in the application. However, a resource utilization overhead is associated with this increase in concurrency. Hence, it is crucial to analyze the profiling data before increasing the number of memory and compute kernels.

We used the Intel OpenCL-to-FPGA profiling tool to analyze the throughput and execution time of each kernel. In order to avoid high resource utilization overhead, we only instantiate multiple versions of the memory and compute kernel for the kernel(s) with the dominant execution time in the application. This rules out kernels commonly used for initialization only once at the beginning of application execution.

Figure 4 shows the speedup from having two *memory* and two *compute* kernels (M2C2) alongside resource utilization overhead. We compare the speedup to the feed-forward design baseline, which indicates that the M2C2 version is much faster than the single work-item baseline model in most applications. Results show an average of 39% speedup over the feed-forward design model baseline with only 31% average increase in logic utilization and a 26% average increase in the number BRAMs used by the implementation. In the case of Pagerank and Back propogation benchmarks, the profiling data from the feed-forward design baseline indicates highly optimized memory operations with high global memory bandwidth utilization. This characteristic of these applications hinders further performance improvement from using multiple *memory* and *compute* kernels. Moreover, it is worth mentioning that analyzing the results from different sections of the experiments indicates no obvious increase/decrease trend for the maximum frequency of the final implementation for different versions of each application.

Further, we explored using four memory and four compute kernels (M4C4) for benchmarks that benefited from the M2C2 transformation; however, except for a performance improvement of ≈20% on HS and BFS and of ≈15% on GC, we did not observe additional performance benefits. Considering the associated resource utilization overhead, M2C2 results the best multiple producers/consumers configuration.

TABLE III: M2C2 Speedup and resource utilization comparison for microbenchmarks

| Benchmark | Baseline Execution Time (ms) | Speedup | Logic Utilization Baseline | Logic Utilization M2C2 | BRAM Baseline | BRAM M2C2 |
|---|---|---|---|---|---|---|
| M AI10 R | 232 | 1.55x | 17.69 | 25.39 | 612 | 892 |
| M AI10 IR | 440 | 1.00x | 17.91 | 24.60 | 817 | 1215 |
| M AI6 for-if R | 10780 | 1.90x | 18.13 | 25.39 | 664 | 892 |
| M AI6 for-if IR | 11500 | 1.84x | 17.71 | 24.35 | 799 | 1161 |

At last, we implemented the M1C2 (one memory and two compute kernels) and M2C1 (two memory and one compute kernels) designs for all the considered benchmarks. As we expected, neither of these versions proved beneficial. Since the main bottleneck in the considered applications are the memory operations, duplicating only the compute kernels in the M1C2 configuration introduces a large number of stalls on the compute kernels. For the M2C1 versions, in order to avoid complex indexing or having separate channels for transferring indices, we have to interleave memory indices read by each memory kernel. This degrades the performance of load units for instructions with regular access patterns.

**Using Vector Variable Types** - Along with using pipes to enable the feed-forward design model, we also tried to improve the memory bandwidth utilization by using vector type operations. Using vector type operations can potentially decrease the number of memory read and write requests and data transfers performed by the kernel. The speedup from using vector type variables is highly dependent on the memory access patterns and memory bandwidth utilization of the OpenCL kernel. For instance, using vector type operations, we were able to improve the throughput of the Floyd-Warshall benchmark by $3\times$ while it degraded the performance of the Maximal Independent Set benchmark significantly. Unfortunately, we could not explore this optimization on all the considered benchmarks due to an internal flaw in the Intel OpenCL-to-FPGA SDK. This flaw results in an internal error while using pipes and vector type memory operations and data transfers. We contacted Intel and they confirmed this compiler flaw.

*2) Experiments on Microbenchmarks:* We designed two sets of automatically generated microbenchmarks to explore the impact of two kernel features on the performance of the feed-forward design model. The first feature is the access pattern of the load instructions, and the second is the control flow divergence among different iterations of the main loop in the single work-item kernel. The first set of microbenchmarks targets memory access patterns. We use two kernels with no control flow divergence across main loop iterations, *eight* load instructions from global memory, and *eighty* arithmetic operations (i.e., arithmetic intensity of 10). These two kernels only differ in the behavior of their load instructions. The first benchmark in this set, called *M AI10 R*, has load instructions with regular memory access patterns, and the second one, called *M AI10 IR*, has load instructions with irregular memory access patterns.

The second set of microbenchmarks targets the presence of control flow divergence within the main loop in a single work-item kernel. To this end, we designed two kernels with the same characteristics as the first set; however, we added an inner loop alongside a conditional statement inside the inner loop to add divergence to the first set of microbenchmarks. To further show the impact of the feed-forward design model on kernels with DLCD, we added a reduction operation inside the inner loop to add data dependencies across different iterations of the inner loop. We also decreased the number of arithmetic operations inside the kernels to increase the control flow divergence impact on the execution time of the kernel. The first microbenchmark, called *M AI6 for-if R*, has load instructions with regular, and another, called *M AI6 for-if IR*, has load instructions with irregular memory access patterns.

Table III shows the impact of the feed-forward design model with two memory kernels and two compute kernels on these sets of microbenchmarks. The results suggest that kernels containing load instructions with regular memory access patterns would often benefit more from the feed-forward design model. This comparison indicates that higher memory contention for irregular load instructions in the feed-forward design with multiple memory kernels, leads to lower memory bandwidth utilization. Moreover, the feed-forward design model improves performance for kernels with control flow divergence and DLCD. The baseline version of these microbenchmarks has a low memory bandwidth utilization due to a more complex control flow and the presence of DLCD. As we recall from Section III, using the feed-forward design model removes the DLCD from the memory kernel and improves the performance of the design on the FPGA. Furthermore, having multiple memory and compute kernels increases the concurrency among memory instructions. These changes will result in significantly higher memory bandwidth utilization and better execution time.

## V. RELATED WORK

Previous works [21] [22] aimed to improve the performance of OpenCL kernels on FPGAs. They explored best practice optimizations from the Intel documentation [17] such as loop unrolling, pipeline replication, and SIMD execution, as well as custom optimizations such as loop coalescing and padding. Our work covers a different code optimization strategy.

Zohouri et al. [13] and Nourian et al. [20] studied several optimization techniques on different applications from Rodinia benchmark suite and finite automata traversal, respectively, while focusing on performance and power consumption. Their analysis confirms the performance portability gap while porting a GPU-optimized OpenCL implementation to FPGA and indicates a critical need for FPGA-specific optimizations to reduce this gap. Krommydas et al. [10] performed a similar analysis on several OpenCL kernels investigating pipeline parallelism on single work-item kernels, manual and compiler vectorization, static coalescing, pipeline replication, and inter-kernel channels. Hassan et al. [11] explored FPGA specific optimizations in their work. Their benchmarks were chosen from irregular OpenCL applications suffering from unpredictable control flows, irregular memory accesses and work imbalance among work-items. In their work, they exploit parallelism at different levels, floating-point optimizations, and data movement overhead across the memory hierarchy.

Several previous works have tried to leverage channels to improve the performance of their implementations by increasing the concurrency among the instructions. Sanaullah et al. [9] proposed an empirically guided optimization framework for OpenCL to FPGA. They leveraged channels to convert a single kernel implementation to multiple kernels, each working as a separate processing element. In their work, they used channels for data communication among kernels. However, their analysis indicates that using channels in their implementation can result in lower performance, mainly due to the data dependency among kernels and the need for synchronizing data paths. Wang et al. [23] leveraged using task kernels and channels to design a multi-kernel approach to reduce the lock overhead. Mainly their work was focused on data partitioning workload. Yang et al. [18] used channels to implement a specific molecular dynamic application.

In a more recent work, Liu et al. [19] proposed a compiler scheme to optimize different types of multi-kernel workloads. They introduced a novel algorithm to find an efficient implementation for each kernel to balance the throughput of a multi-kernel design. Additionally, they explored bitstream splitting to separate multiple kernels into more than one bitstream to enable more optimizations for individual kernels.

## VI. CONCLUSION

In this work, we proposed a systematic code transformation method to apply the feed-forward design model to existing OpenCL kernels. We evaluated the benefits and limitations of our method, as well as its applicability in the presence of different classes of loop-carried dependencies. We performed experiments on a set of applications from broadly used benchmark suites. We leveraged the Intel OpenCL-to-FPGA SDK for compiling and profiling the kernels before and after the transformation. We analyzed the effect of the memory access and compute patterns of the kernels on the performance of the resulting code. Our experiments show speedups up to $86\times$ at a modest resource utilization cost.

## REFERENCES

[1] "Top 500 list," iD: doc:605059a18f08377acf72d2ac; M1: Web Page. [Online]. Available: https://www.top500.org/ https://www.top500.org/

[2] Z. Wang, B. He, W. Zhang, and S. Jiang, "A performance analysis framework for optimizing opencl applications on fpgas," in *2016 IEEE International Symposium on High Performance Computer Architecture (HPCA)*, 2016, pp. 114–125.

[3] "Microsoft azure fpga." [Online]. Available: https://docs.microsoft.com/en-us/azure/machine-learning/how-to-deploy-fpga-web-service

[4] "Amazon ec2 f1," iD: doc:605062b38f08a4e6d25943f1; M1: Web Page. [Online]. Available: https://aws.amazon.com/ec2/instance-types/f1/ https://aws.amazon.com/ec2/instance-types/f1/ https://aws.amazon.com/ec2/instance-types/f1/

[5] "Opencl." [Online]. Available: https://www.khronos.org/opencl/

[6] "Opencl memory model." [Online]. Available: https://www.xilinx.com/html_docs/xilinx2017_4/sdaccel_doc

[7] Intel, "Intel fpga sdk for opencl pro edition: Programming guide." [Online]. Available: https://tinyurl.com/2p87v84n

[8] A. Sanaullah and M. C. Herbordt, "Unlocking performance-programmability by penetrating the intel fpga opencl toolflow," 2018, pp. 1–8, iD: doc:605061558f086a39be7d878c; ID: 1; M1: Conference Proceedings.

[9] A. Sanaullah, R. Patel, and M. Herbordt, "An empirically guided optimization framework for fpga opencl," in *2018 International Conference on Field-Programmable Technology (FPT)*, Dec 2018, pp. 46–53.

[10] K. Krommydas, A. E. Helal, A. Verma, and W. Feng, "Bridging the performance-programmability gap for fpgas via opencl: A case study with opendwarfs," in *- 2016 IEEE 24th Annual International Symposium on Field-Programmable Custom Computing Machines (FCCM)*, 2016, p. 198, iD: 1.

[11] M. W. Hassan, A. E. Helal, P. M. Athanas, W. Feng, and Y. Y. Hanafy, "Exploring fpga-specific optimizations for irregular opencl applications," in *- 2018 International Conference on ReConFigurable Computing and FPGAs (ReConFig)*, 2018, pp. 1–8, iD: 1.

[12] M. E. Zarch, R. Neff, and M. Becchi, "Exploring thread coarsening on fpga," in *2021 IEEE 28th International Conference on High Performance Computing, Data, and Analytics (HiPC)*, 2021, pp. 436–441.

[13] H. R. Zohouri, N. Maruyama, A. Smith, M. Matsuda, and S. Matsuoka, "Evaluating and optimizing opencl kernels for high performance computing with fpgas," in *- SC '16: Proceedings of the International Conference for High Performance Computing, Networking, Storage and Analysis*, 2016, pp. 409–420, iD: 1.

[14] H. R. Zohouri and S. Matsuoka, "The memory controller wall: Benchmarking the intel fpga sdk for opencl memory interface," 11 2019, pp. 11–18.

[15] S. Che, M. Boyer, J. Meng, D. Tarjan, J. W. Sheaffer, S. Lee, and K. Skadron, "Rodinia: A benchmark suite for heterogeneous computing," in *- 2009 IEEE International Symposium on Workload Characterization (IISWC)*, 2009, pp. 44–54, iD: 1.

[16] S. Che, B. M. Beckmann, S. K. Reinhardt, and K. Skadron, "Pannotia: Understanding irregular gpgpu graph applications," in *- 2013 IEEE International Symposium on Workload Characterization (IISWC)*, 2013, pp. 185–195, iD: 1.

[17] "Intel fpga sdk for opencl best practice guide."

[18] Z. Wang, B. He, and W. Zhang, "A study of data partitioning on opencl-based fpgas," in *2015 25th International Conference on Field Programmable Logic and Applications (FPL)*, 2015, pp. 1–8.

[19] J. Liu, A.-A. Kafi, X. Shen, and H. Zhou, *MKPipe: A Compiler Framework for Optimizing Multi-Kernel Workloads in OpenCL for FPGA*, ser. Proceedings of the 34th ACM International Conference on Supercomputing. New York, NY, USA: Association for Computing Machinery, 2020. [Online]. Available: https://doi.org/10.1145/3392717.3392757

[20] M. Nourian, M. E. Zarch, and M. Becchi, "Optimizing complex opencl code for fpga: A case study on finite automata traversal," in *2020 IEEE 26th International Conference on Parallel and Distributed Systems (ICPADS)*, 2020, pp. 518–527.

[21] A. Sanaullah, R. Patel, and M. Herbordt, "An empirically guided optimization framework for fpga opencl," in *2018 International Conference on Field-Programmable Technology (FPT)*, 2018, pp. 46–53.

[22] H. R. Zohouri, A. Podobas, and S. Matsuoka, "High-performance high-order stencil computation on fpgas using opencl," in *2018 IEEE International Parallel and Distributed Processing Symposium Workshops (IPDPSW)*, 2018, pp. 123–130.

[23] C. Yang, J. Sheng, R. Patel, A. Sanaullah, V. Sachdeva, and M. C. Herbordt, "Opencl for hpc with fpgas: Case study in molecular electrostatics," in *2017 IEEE High Performance Extreme Computing Conference (HPEC)*, 2017, pp. 1–8.